\documentclass[amsmath,floats,floatfix,nofootinbib,notitlepage,prd,twocolumn]{revtex4-1}
\DeclareMathAlphabet{\mathpzc}{OT1}{pzc}{m}{it}
 \usepackage{hyperref}
\usepackage{graphicx}
\usepackage{amsmath}
\usepackage{amsfonts}
\usepackage{calligra}
\usepackage{comment}
\usepackage{mathrsfs}
\usepackage{bm}
\usepackage{tensind}
\usepackage{aas_macros}
\usepackage{subfigure}
\usepackage{color}
\newcommand{\pd}{\partial}

\begin{document}

\title{Plasma Waves and Jets from Moving Conductors}
%\title{Conductor-launched Plasma Waves and Jets}
%\title{Plasma Waves and Jets from Conductors in Motion}
\author{Samuel E. Gralla\footnote{{\tt sgralla@email.arizona.edu}} and Peter~Zimmerman\footnote{{\tt peterzimmerman@email.arizona.edu}}}
\affiliation{Department of Physics, University of Arizona}
\date{\today}

\begin{abstract}
We consider force-free plasma waves launched by the motion of conducting material through a magnetic field.  We develop a spacetime-covariant formalism for perturbations of a uniform magnetic field and show how the transverse motion of a conducting fluid acts as a source.  We show that fast-mode waves are sourced by the compressibility of the fluid, with incompressible fluids launching a pure-Alfv\'en outflow.  Remarkably, this outflow can be written down in closed form for an \textit{arbitrary} time-dependent, nonaxisymmetric incompressible flow.  The instantaneous flow velocity is imprinted on the magnetic field and transmitted away at the speed of light, carrying detailed information about the conducting source at the time of emission.  These results can be applied to transients in pulsar outflows and to jets from neutron stars orbiting in the magnetosphere of another compact object.  We discuss jets from moving conductors in some detail.   %As applications, we give a detailed description of the jets from rigidly moving and rotating conductors and write down the analytic solution for dual jets from a binary. \SG{if we can!}
\end{abstract}

\maketitle
\section{Introduction}

In 1831 Faraday discovered that a conductor moving through a magnetic field can power an electric circuit \cite{faraday1832}, a phenomenon now known as unipolar induction.  In 1967 a \textit{yottawatt} unipolar inductor was discovered in nature, the first known pulsar \cite{hewish-etal1968}.  Pulsars are rotating, magnetized neutron stars surrounded by a diffuse plasma magnetosphere \cite{gold1968,pacini1968,goldreich-julian1969,ruderman-sutherland1975}.  The rotation induces a voltage that drives current through the magnetosphere, carrying energy away in a manner precisely analogous to Faraday's original device.

Laboratory unipolar inductors have been studied extensively as efficient electrical generators for specialized tasks requiring low internal resistance, and the basic theory is very well understood.  The theory of unipolar induction in plasma physics is comparatively underdeveloped, especially in the low-density (force-free) regime relevant to pulsars.  The many remaining puzzles of pulsar phenomenology, along with the likely role of magnetized neutron stars in other mysterious events such as gamma-ray bursts, motivate a systematic treatment of the basic mechanism of unipolar induction in force-free plasma. 

We consider the idealized setup of a uniform magnetic field perturbed by a conducting fluid flowing orthogonally in a plane.  We give a general treatment of linearized force-free perturbations and identify the two basic modes of propagation, the fast and Alfv\'en modes.  We show that the fast mode couples to the expansion of the fluid, so that incompressible fluids launch only Alfv\'en waves.  Remarkably, we are able to write the Alfv\'en outflow in closed form for an \textit{arbitrary} time-dependent, nonaxisymmetric incompressible flow.  The instantaneous flow velocity is imprinted on the magnetic field and transmitted away at the speed of light.  The electromagnetic field is a``light dart'' \cite{gralla-jacobson2015}, part of a large class of analytic solutions with null four-current  \cite{bateman1913,bateman1923,michel1973mon,menon-dermer2007,lyutikov2011collapse,brennan-gralla-jacobson2013,brennan-gralla2014,gralla-jacobson2015}. %\SG{cite NHEK work'?}

These results inform the discussion of pulsar outflows and are promising for understanding transients.   For a magnetic field emerging from a conducting star, we can apply the results to a local patch where the surface and field are both approximately uniform, applying a local boost to make the flow orthogonal to the field.  %\footnote{If the field does not emerge orthogonally, one can make this approximately so by boosting along the field lines.}   
  This illustrates the pure-Alfv\'en nature of pulsar outflows from highly incompressible neutron star crusts, and more importantly points to compressibility as a source of fast-mode waves.  Heyl and Hernquist \cite{heyl-hernquist1998a,heyl-hernquist1998b,heyl-hernquist1997,heyl-hernquist2003,heyl2005,heyl2006} have shown that, for high-field pulsars, fast-mode waves will steepen and shock due to the effects of quantum electrodynamics, and that the dissipation into electron-positron pairs could power an X-ray flare.  One criticism of the model is the lack of a mechanism to generate the required fast-mode disturbance.  Here we show that any violation of incompressibility will launch fast-mode waves, suggesting that a shearing or breaking of the crust would create a large-amplitude fast-mode disturbance.% that could potentially power a flare.  %We are currently investigating the details.

\begin{figure*}
\centering
\subfigure[\ rotating]{\label{fig:rotjet}\includegraphics[height=0.48\textwidth]{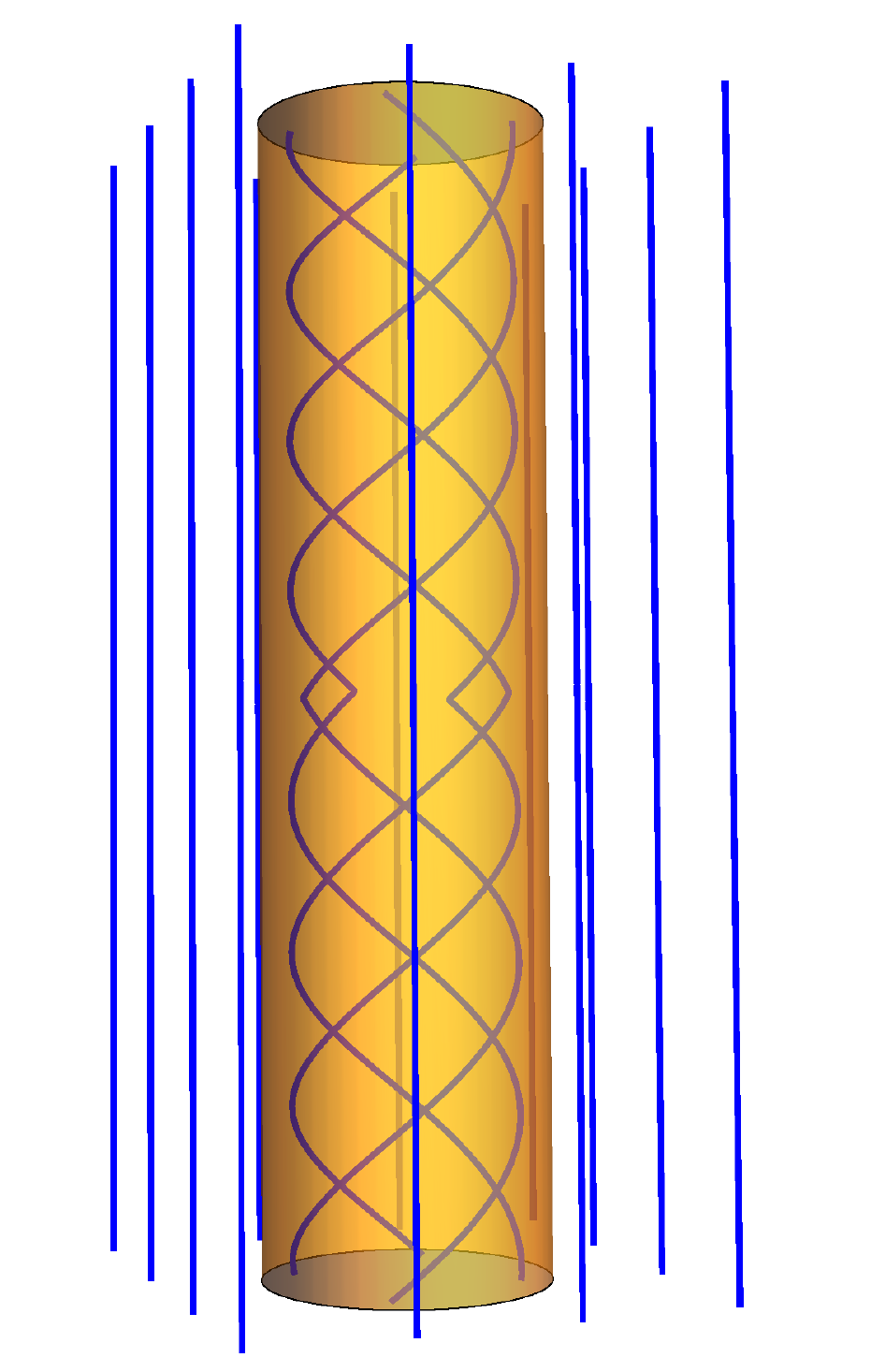} \ \ \ \ \ \ \ \ \ \ \ \  }
\subfigure[\ moving]{\label{fig:movejet}\includegraphics[height=0.48\textwidth]{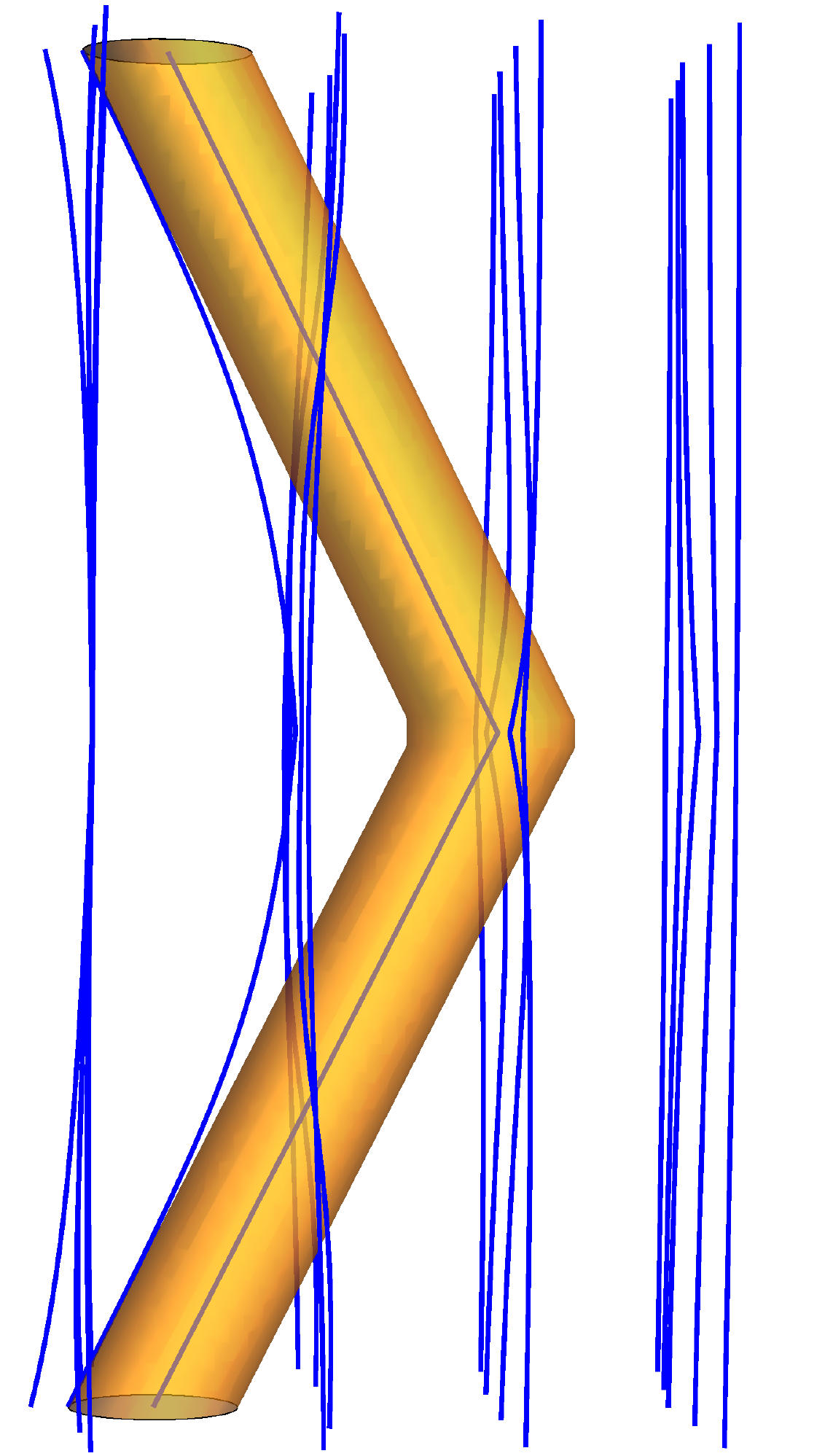}\ \ \ \ \ \ \ \ \ }
\subfigure[\ binary]{\label{fig:binaryjet}\includegraphics[height=0.49\textwidth]{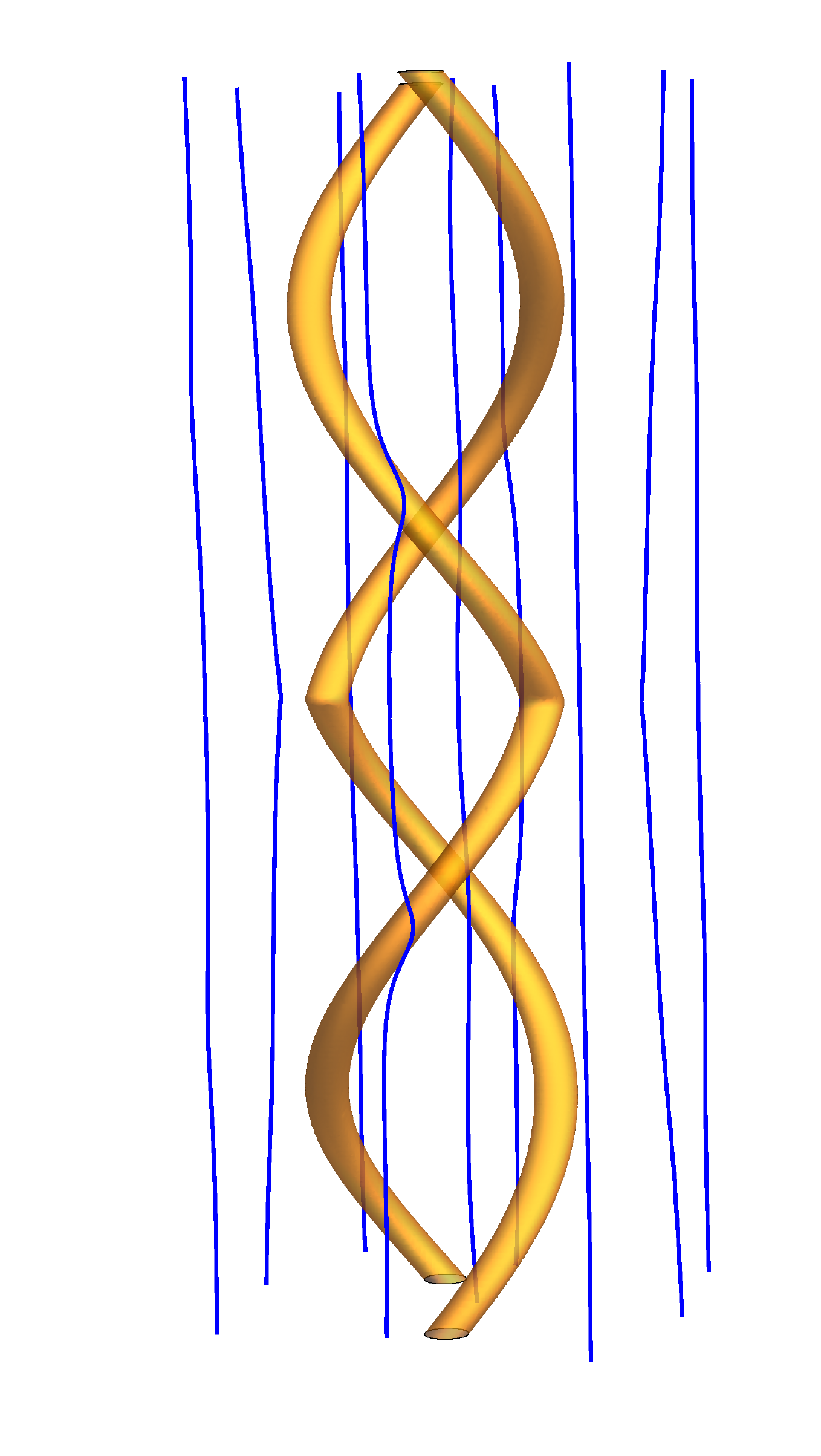}}
\caption{Analytic solutions for jets from thin conducting disks moving in a magnetic field.  The current sheet surrounding the jet is in orange and the magnetic field lines are in blue.  From left to right, a disk rotating counter-clockwise, a disk moving to the right, and a pair of disks rigidly orbiting in a counter-clockwise direction.   The disks move in the plane pictured mid-way up the diagram, and the field lines are seeded at equally spaced intervals in this plane.   In the non-rotating cases (b) and (c), the magnetic field in the jet points along it and the current is carried entirely on the edge. }
\label{fig:jets}
\end{figure*}

 A second application is to the motion of a conducting object through a magnetic field that varies on larger spatial scales.  This process has been extensively studied in the context of the terrestrial and Jovian  magnetospheres \cite{drell-foley-ruderman1965,goldreich-lynden-bell1969}, but we are unaware of a previous detailed analysis for the relativistic force-free plasmas of compact object magnetospheres.  To fill this gap, we apply our formalism to thin, rigid conductors of arbitrary shape moving orthogonally to a uniform field with arbitrary time-dependent non-relativistic velocity.  The power radiated scales as
\begin{align}\label{scaling}
P \propto B^2 v^2 L^2
\end{align}
for a body of typical size $L$ moving through a magnetic field $B$ with velocity $v$.  The coefficient depends on the details of the body shape [e.g. Eq.~\eqref{Pline} below].  While the basic result \eqref{scaling} is well-known, our treatment reveals much detail that we have not seen previously discussed.  In particular, we show that 1) Eq.~\eqref{scaling} holds for arbitrary motion of the body (transverse to the field), i.e., there are no corrections due to acceleration; 2) the shape of the jet is fixed by the profile of the conductor at the retarded time; 3) the magnetic field in the jet always points along it; 4) the current is entirely carried on a thin layer on the edge of the jet.  We provide complete analytic solutions for disk-shaped conductors -- see Sec.~\ref{sec:disk} and Fig.~\ref{fig:jets}.

%\SG{fix}
%\begin{align}\label{GiveMeThePower}
%P = 2 B^2 A_{\perp} v_{\perp}^2\sqrt{1-v_{\perp}^2}, 
%\end{align}
%where $B$ is the magnetic field strength, $v_{\perp}$ is the velocity perpendicular to the field, and $A_{\perp}$ is the area of the conductor (in its rest frame\footnote{The factor of $1/\gamma_{\perp}=\sqrt{1-v_{\perp}^2}$ arises because the power is proportional to the cross-sectional area in the magnetic field frame.}) projected along the field.  We provide explicit global analytic solutions for disk-shaped conductors in a variety of states of motion -- see Fig.~\ref{fig:jets}.

These results apply to real conductors on scales far enough from the conductor that it can be considered thin, but near enough that the field can be considered uniform.  The analysis predicts jets from neutron stars orbiting near an active supermassive black hole (where there is expected to be a force-free plasma \cite{blandford-znajek1977}), a system which would also emit gravitational waves detectable by eLISA \cite{eLISA}.  It could also be relevant for electromagnetic precursor signals from neutron star mergers, the likely progenitor of gamma-ray bursts.  The current sheet surrounding the jet provides a possible region for particle acceleration and electromagnetic radiation.

Finally, we mention that following the seminal work of Blandford and Znajek \cite{blandford-znajek1977}, it has been recognized that black holes can act as unipolar inductors and drive a plasma circuit in much the same way as conductors.  While the case of a rotating black hole has been studied extensively for four decades, the theory of moving black holes is much more in its infancy %\cite{palenzuela-etal2010,palenzuela-lehner-liebling2010,neilsen-etal2011,lyutikov2011unipolar,mcwilliams-levin2011,lai2012,alic-etal2012,paschalidis-etienne-shapiro2013,dorazio-levin2013,morozova-rezzolla-ahmedov2014,penna2015}.  
\cite{neilsen-etal2011,lyutikov2011unipolar,morozova-rezzolla-ahmedov2014,penna2015}, with the main result being the $B^2 v^2$ scaling of the power radiated \cite{neilsen-etal2011}, in agreement with the conductor case.  Our detailed results for conductors provide a foil for studying the black hole case in more detail.  Combined with recent advances in the theory of force-free perturbations in curved spacetime \cite{yang-zhang2014,yang-zhang-lehner2015,yang-zhang2016,zhang-mcwilliams-pfeiffer2015}, we are optimistic that it will soon become possible to perform detailed comparisons between conductor-driven and black-hole-driven plasma circuits, analogously to recent work \cite{gralla-lupsasca-rodriguez2016} in the rotating case.  

The paper is organized as follows.  In Sec.~\ref{sec:linear} we consider force-free perturbations of a uniform magnetic field, and in Sec.~\ref{sec:sources} we show how conducting fluid launches plasma waves.  In Sec.~\ref{sec:jets} we study jets from isolated conductors, and in Sec.~\ref{sec:disk} we explicitly study disk-shaped conductors.  We use Heavside-Lorentz units with $c=1$. Our flat metric has signature $(-1,1,1,1)$ and we use the orientation $\epsilon_{txyz}=1$.  Lowercase Latin indices run over $t$ and $z$, while uppercase Latin indices run over the transverse directions ($x$ and $y$ in Cartesian coordinates).  Greek indices run over all spacetime coordinates. 

\section{Linearized force-free equations}\label{sec:linear}

Any force-free field, or more generally any degenerate, closed two-form, may be expressed in terms of a pair of scalar potentials $\theta_1$ and $\theta_2$ as \cite{carter1979,uchida1997general,gralla-jacobson2014}
\begin{align}
F= d  \theta_1  \wedge d\theta_2.
\end{align} 
We perturb a uniform field $F=B_0 dx \wedge dy$, setting $B_0=1$ for notational convenience.  (Factors of $B_0$ can be restored on dimensional grounds.)  The potentials take the form
\begin{align*}
\theta_1 &= x + \alpha \\ 
\theta_2 &= y + \beta ,
\end{align*} 
where $\alpha$ and $\beta$ are considered small.  (We keep to linear order.)  The perturbed field strength takes the form 
  \begin{align} 
  F = dx\wedge dy + d\alpha \wedge dy + dx \wedge d\beta, 
  \end{align}
  and notably lacks a $dt \wedge dz$ component due to its degeneracy.  We impose the force-free condition
\begin{align} 
F_{\alpha \beta}  j^\alpha = 0 , 
\end{align}
  where $ j^\alpha=\pd_\beta F^{\alpha \beta}$  is the charge-current four-vector.
%current vector $ j^\alpha  = \tfrac{1}{3!} \epsilon^{\alpha\beta\gamma\delta}J_{\alpha\beta\gamma}$, and $J_{\alpha\beta\gamma} = (\pd_\nu F^{\mu\nu}) \epsilon_{\mu\alpha\beta\gamma}$ are the tensor components of the current three-form $J = d \star F$.  
The linearized force-free equations are then 
\begin{align} 
 - \pd_t^2 \alpha +  \pd_x^2  \alpha + \pd_z^2 \alpha +  \pd_x \pd_y \beta  &=0\\
 -  \pd_t^2 \beta +  \pd_y^2 \beta + \pd_z^2 \beta + \pd_x \pd_y \alpha &=0.
\end{align}
Note the equations are symmetric under the simultaneous interchange of $\alpha \leftrightarrow \beta$ and $x\leftrightarrow y$.  The second derivative coupling can be eliminated\footnote{We could alternatively decouple the equations by taking derivatives, following \cite{yang-zhang2014,yang-zhang-lehner2015,yang-zhang2016}.  The present approach has the advantage that the field strength is given directly by derivatives of the potentials, rather than requiring integration.} by introducing a pair of ``Hertz'' potentials  $\Psi$ and $\Phi$ by
\begin{subequations}\label{eqs:CR-HPs}
\begin{align}
\pd_x \Phi + \pd_y \Psi &= \alpha, \\    
\pd_y \Phi - \pd_x \Psi &=   \beta,
\end{align} 
\end{subequations}
which always exist by the Helmholtz theorem.  The force-free equations then become
 \begin{subequations}\label{eqs:CR-WHPs}
\begin{align} 
 \pd_x \Box  \Phi & +  \pd_y \Box_2 \Psi  = 0, \\
 \pd_y \Box \Phi  & - \pd_x \Box_2\Psi = 0 ,
\end{align}
\end{subequations}
where $\Box :=-\partial_t^2+\partial_x^2 + \partial_y^2 + \partial_z^2$ and $\Box_2 :=-\partial_t^2+\partial_z^2$.  These are the Cauchy-Riemann equations for $\Box_2 \Psi + i  \, \Box \Phi$, so the general solution is
\begin{subequations}\label{eqs:unfixed}
\begin{align}
\Box_2 \Psi = \textrm{Re}[f(x+i y,t,z)] \label{eq:box2psi=Ref} \\
\Box \Phi = \textrm{Im}[f(x+i y,t,z)], \label{eq:boxPhi=Imf}
\end{align}
\end{subequations}
where $f$ is holomorphic in its complex argument.  The field strength is given in terms of the Hertz potentials by
\begin{subequations}\label{eqs:F-comps}
\begin{align}
F_{ab} &=0, \\
F_{aB} &= \pd_a \pd_B \Psi + \epsilon^{C}_{\  B} \pd_a \pd_C \Phi, \\
F_{AB} &= \epsilon_{AB}\left( 1 + \triangle \Phi \right), \label{FAB}
\end{align}
\end{subequations}
where $a,b,c...$ range over $t$ and $z$, $A,B,C...$ range over $x$ and $y$, $\epsilon_{xy}=1$ is antisymmetric, and $\triangle=\pd_x^2+\pd_y^2$. (More formally, we perform a $2+2$ decomposition into the Lorentzian $tz$ space and the transverse Euclidean $xy$ space, and $\bm{\epsilon}$ and $\triangle$ are the transverse volume element and Laplacian.)  The gauge freedom in this representation is
\begin{subequations}\label{eqs:gauge}
\begin{align}
\Psi & \rightarrow \Psi + \textrm{Re}[g(x+ i y,t,z)]+h(x,y) \\
\Phi & \rightarrow \Phi + \textrm{Im}[g(x+iy,t,z)],
\end{align}
\end{subequations}
for real functions $h$ and complex functions $g$ (i.e., $g$ is holomorphic in the $xy$ complex plane).  By solving $\Box_2 g=f$ we can always eliminate $f$ from Eqs.~\eqref{eqs:unfixed}, leaving the simpler equations
\begin{align}\label{eq:FF}
\Box_2 \Psi = 0, \quad \Box \Phi = 0.
\end{align}
The remaining gauge freedom is Eqs.\eqref{eqs:gauge} with $\Box_2g=0$.  We will work exclusively in this class of gauges, taking Eqs.~\eqref{eq:FF} as the fundamental equations.  
 Following standard terminology, we will refer to $\Psi$ and $\Phi$ as the Alfv\'en and fast modes, respectively. 

The general solution for $\Psi$ is $\Psi=\Psi_L(t + z, \, x,y) + \Psi_R( t - z, \, x,y)+Ct + Dz+E$, where $C,D,E$ are constants.  The latter three terms can be removed by a final gauge transformation $g=Ct + Dz+E$.  In this case the complete field equations are
\begin{subequations}\label{eq:FF2}
\begin{align}
\Psi & = \Psi_L(t + z, \, x,y) + \Psi_R( t - z, \, x,y) \label{FFPsi} \\
\Box \Phi & = 0. \label{FFPhi}
\end{align}
\end{subequations}
Explicitly, the remaining gauge freedom is 
\begin{subequations}\label{eqs:residual_gauge}
\begin{align}
\Psi_L & \rightarrow \Psi_L + \textrm{Re}[g_L] + h_L(x,y) \label{PsiLgauge} \\
\Psi_R & \rightarrow \Psi_R + \textrm{Re}[g_R] + h_R(x,y)  \label{PsiRgauge}\\
\Phi & \rightarrow \Phi + \textrm{Im}[g_L] + \textrm{Im}[g_R] + Ct + Dz + E \label{Phigauge}
\end{align}
\end{subequations}
where $h_L$ and $h_R$ are real functions, $C,D,E$ are constants, and $g_L$ and $g_R$ are complex functions of the form
\begin{align}\label{gLR}
g_L = g_L(x+ i y,t+z), \quad g_R=g_R(x+iy,t-z).
\end{align}
Here the notation $x+iy$ indicates that each is holomorphic in the $xy$ complex plane.

The four-current $j$ associated with \eqref{eqs:F-comps} is $j = \pd^a (\triangle \Psi) \pd_a$.  In terms of the left and right moving modes \eqref{FFPsi}, we have
\begin{align}\label{j}
j = -\triangle \dot{\Psi}_L (\partial_t - \partial_z) - \triangle \dot{\Psi}_R (\partial_t + \partial_z),
\end{align}
where an overdot is a derivative with respect to $t$.  Thus only the Alfv\'en mode carries current.  Notice that the three-current is always in the $z$ direction, along the background field lines.

\subsection{Pure-Alfv\'en perturbations}\label{sec:pure-alfven}

If there is a gauge where $\Phi=0$ then we say that the perturbation is pure-Alfv\'en.  From Eqs.~\eqref{eqs:F-comps} we see that pure-Alfv\'en solutions are transverse: they do not perturb the $z$ magnetic field $B_z=F_{AB}$.  In fact the converse is also true: if $B_z$ is unperturbed, then the perturbation is pure-Alfv\'en,
\begin{align}
\delta B_z = 0 \quad \leftrightarrow \quad \textrm{pure-Alfv\'en}.
\end{align}
To see this, note from \eqref{eqs:F-comps} that $\delta B_z=\triangle \Phi$.  If $\triangle \Phi$ vanishes, then by the field equation $\Box \Phi=0$ we also have $\Box_2\Phi=0$.  But the gauge freedom \eqref{Phigauge} is the general solution to the equations $\triangle \Phi=\Box_2\Phi=0$, so we may set $\Phi=0$ by a gauge transformation.

In general there can be both left and right moving Alfv\'en modes, Eq.~\eqref{FFPsi}.  However, if one mode vanishes then the pure-Alfv\'en solution has special properties.   In the right-moving case ($\Psi_L=0$) we introduce a new variable
\begin{align}\label{chi}
\chi = -\partial_t \Psi_R(t-z,x,y), 
\end{align}
which makes the field strength \eqref{eqs:F-comps} take the elegant form,
\begin{align}\label{light-dart}
F = dx \wedge dy + d \chi \wedge d(t-z).
\end{align}
This is in fact an \textit{exact} solution of force-free electrodynamics (not merely linearized), the ``light dart'' discussed in \cite{gralla-jacobson2015}.  It is a time-dependent, nonaxisymmetric exact solution parameterized by a free function $\chi(t-z,x,y)$ of three variables.  The Poynting flux (energy flux per unit time per unit area) in the $z$-direction is given by 
\begin{align}\label{fluxdart}
 \mathcal{F}_{\rm Poynting} =  \pd_A \chi \, \pd^A \chi.
\end{align}
%\SG{The linear and angular momenta are kind of ugly I think.}\PZ{They're  the light dart fluxes
%\begin{align*}
%\mathcal{F}_{P_z} &= - \frac12 B_0^2 + \pd_A \chi \, \pd^A \chi \\
%\mathcal{F}_{L_z} &= x^A \pd_A \chi,
%\end{align*}}
Finally, the current is
\begin{align}\label{jdart}
j = \triangle \chi \left( \partial_t + \partial_z \right),
\end{align}
which is null.

If we instead consider a purely left-moving wave ($\Psi_R=0$), then we let $\chi =- \partial_t \Psi_L(t+z,x,y)$ and Eq.~\eqref{light-dart} becomes $F = dx \wedge dy + d \chi \wedge d(t+z)$, while Eq.~\eqref{fluxdart} becomes $\mathcal{F}_{\rm Poynting} = - \pd_A \chi \, \pd^A \chi$ and Eq.~\eqref{jdart} becomes $j = \triangle \chi \left( \partial_t - \partial_z \right)$.  %This is also an exact solution, but the superposition of left- and right-moving solutions is only valid in linearized theory. 

%\subsection{Stationary configurations}

%\SG{Do we need this?  It kind of breaks the flow. Put in an appendix?}Suppose the fields are stationary, i.e., $\mathcal{L}_{\pd_t} F_{\alpha \beta}=0$ (or $\pd_t F_{\alpha \beta}=0$ in Minkowski coordinates).  From $\partial_t F_{AB}=0$ we learn that $\Phi=\Phi_0(x,y,z)+\Phi_H(x,y,z,t)$ where $\triangle \Phi_H=0$.  The field equation $\Box \Phi=0$ then implies that $\nabla^2 \Phi_0=0$ and $\Box_2\Phi_H=0$.  But then we may use the residual gauge freedom \eqref{eqs:residual_gauge} to set $\Phi_H=0$.  Thus $\Phi=\Phi_0(x,y,z)$.  Then from $\partial_t F_{aB}=0$ and \eqref{FFPsi} we learn $\Psi=(t+z)\hat{\Psi}_L(x,y) + (t-z)\hat{\Psi}_R(x,y) + h(x,y)$.  (The hats distinguish from the general case \eqref{FFPsi}.) The last term can be removed by the remaining gauge freedom in \eqref{eqs:residual_gauge}.  Thus we find that for stationary solutions we may always write
%\begin{subequations}\label{eqs:stats}
%\begin{align}
%\Psi &= (t+z) \hat{\Psi}_L(x,y) + (t-z) \hat{\Psi}_R(x,y) \label{eq:Psistat} \\
%\Phi &= \Phi(x,y,z), \quad \nabla^2 \Phi=0. \label{eq:Phistat}
%\end{align}
%\end{subequations}
%Conversely, if Eqs.~\eqref{eqs:stats} are satisfied, then the linearized force-free equations are satisfied.  

\section{Conducting sources}\label{sec:sources}

In the previous section we developed the theory of force-free fields linearized off of a uniform magnetic field, showing the presence of the two basic modes of propagation.  The analysis was agnostic as to the cause of such waves, i.e., it was carried out in the absence of ``sources''.  In the context of force-free electrodynamics, conductors act as sources by imposing a boundary condition on the external force-free fields.  If $u^\alpha$ is the four-velocity field of the surface of a perfect conductor and $S$ is its three-dimensional timelike worldvolume, then the condition is
\begin{align}\label{conductingBC}
\left[F \cdot u \right]_{S^+} =0,
\end{align}
where $[\dots]_{S^+}$ denotes pullback\footnote{The pullback of a form is simply the form regarded as a form on $S$.  Algebraically, if the surface is specified by $\sigma=0$ for some scalar field $\sigma$, the pullback entails setting $\sigma=0$ and $d\sigma=0$ in the coordinate-basis expression for the form.} to $S$ from the outside.  In general the pullback of the Maxwell field $F$ across any boundary layer must be continuous to avoid magnetic monopole surface charge and current \cite{gralla-jacobson2014}.  Eq.~\eqref{conductingBC} results from demanding that the rest-frame electric field $F\cdot u$ vanish inside the conductor (definition of a perfect conductor), so that that pullback of this quantity from the outside must vanish as well.  The condition \eqref{conductingBC} allows the existence of electric surface charge and current, which is proportional to the jump in the pullback of the dual of $F$ \cite{gralla-jacobson2015}.

Now consider a conducting fluid moving in the $xy$ plane with four-velocity field $u^\mu=dx^\mu/d\tau$.  We express the flow in terms of the boost factor $\gamma$ and the three-velocity $v^A$,
\begin{align}
u  = \gamma (x^A,t) \left[ \partial_t + v^A(x^A,t) \partial_A \right].
\end{align} 
%and also introduce the two-dimensional expansion $\theta$ and vorticity $\omega$,
%\begin{align}
%\theta = \pd_A v^A, \quad \omega =\scalebox{1.25}[1.25]{$\epsilon$}^{AB} \pd_A v_B.
%\end{align}
The boundary condition \eqref{conductingBC} then becomes
\begin{align}\label{eq:pecker}
  \pd_t \left( \pd_B \Psi +  \scalebox{1.25}[1.25]{$\epsilon$}^{C}_{\ B} \pd_C \Phi \right) =  \scalebox{1.25}[1.25]{ $\epsilon$}_{BC} v^C(1 + \triangle \Phi ),
\end{align}
holding at $z=0$.  Acting with $\scalebox{1.25}[1.25]{ $\epsilon$}^{AC} \pd_C$ yields
%and $\partial_A$ yields (respectively) \SG{$\gamma$ added -- check}
\begin{align}\label{almost}
\partial_t \triangle \Phi & = - \pd_A v^A (1+\triangle \Phi) -v^A \pd_A \triangle \Phi 
\end{align}
Introducing the Lagrangian derivative $d/d\tau$ and the two-dimensional expansion $\theta$,
\begin{align}
\frac{d}{d\tau} & = u^\mu \pd_\mu = \gamma(\partial_t+v^A \pd_A) \\
\theta & = \nabla_A v^A,
\end{align}
Eq.~\eqref{almost} becomes 
\begin{align}\label{nugget}
\frac{d}{d\tau} \triangle \Phi & = - \gamma \theta (1+\triangle \Phi),
\end{align}
still holding at $z=0$.  Recall that $\triangle \Phi$ is the perturbation in the $z$ magnetic field.  

Eq.~\eqref{nugget} can be solved for $\triangle \Phi$ on $z=0$ by introducing Lagrangian coordinates $(\tau,\hat{x}^A)$ (defined so that $\hat{x}^A$ is constant along the flow) and using the method of integrating factors.  One may then obtain $\Phi$ by inverting the Laplacian and extend off of $z=0$ by solving $\Box \Phi=0$. 
%\begin{align}
%\triangle \Phi & =  A(\hat{x}^A) \,  \exp\left[ - \int_0^\tau \gamma \, \theta \, d \tau' \right],  \label{triPhi}
%\end{align}
%where $A(x^A)=\triangle \Phi(t=0,x^A)$ is the initial data.  We may obtain $\Phi$ itself by inverting the Laplacian and then extending off of $z=0$ by solving $\Box \Phi=0$.
With $\Phi$ is determined, $\Psi$ is given on $z=0$ by Eq.~\eqref{eq:pecker}.
%\begin{align}\label{fun}
 %\pd_t \pd_B \Psi =  \scalebox{1.25}[1.25]{ $\epsilon$}_{BC} v^C(1 + \triangle \Phi ) - \scalebox{1.25}[1.25]{$\epsilon$}^{C}_{\ B}  \pd_C\,  \partial_t \Phi.
%\end{align}
%\PZ{We already have this equation above. It's  redundant to repeat it here. }
If the wave is assumed purely right-moving or left-moving then Eq.~\eqref{FFPsi} fixes $\pd_t \pd_B \Psi$ everywhere.  It also fixes $\pd_z \pd_B \Psi$, which determines the perturbation \eqref{eqs:F-comps} everywhere.

\subsection{Incompressible flow}\label{sec:incompressible}

Eq.~\eqref{nugget} shows how changes in the longitudinal component of field, $\delta B_z=\triangle \Phi$, are sourced by the expansion $\theta$ of the flow.  If the expansion vanishes (incompressible flow), then $B_z$ is conserved along the flow,
\begin{align}
\frac{d}{d\tau} \triangle \Phi = 0.
\end{align}
In particular, if $\delta B_z=\triangle \Phi$ vanishes initially, then it vanishes for all time on $z=0$.   As discussed in Sec.~\ref{sec:pure-alfven}, we can then set $\Phi=0$ entirely on $z=0$ by a gauge choice.  If the conducting fluid started to move at some finite time in the past, before which the uniform field was unperturbed, it follows that $\Phi=0$ everywhere.\footnote{This setup constitutes an initial boundary value problem with $\Phi=\partial_t \Phi=0$ everywhere at an initial time, $\Phi=0$ on $z=0$ for all time, and $\Box \Phi=0$ on $z\neq 0$ for all time.  We may use standard energy estimate techniques to show that the solution $\Phi=0$ is unique.  In particular, we may modify the version given by Wald \cite{wald-book1984} (see Fig. 10.1 therein) by cutting off the integration volume $K$ at $z=0$.  This generates a new flux integral on $z=0$ in Eq.~(10.1.10), but the integral vanishes because $\Phi=0$ on $z=0$.  The rest of the proof proceeds unmodified.  If we know that $\Phi=0$ only on a portion of $z=0$ (as needed for the conductors of finite extent treated later), we may suitably modify $K$ and again conclude that $\Phi=0$ is the unique solution.} In this sense, incompressible flows do not launch fast waves.

Incompressible flows do launch Alfv\'en waves.  For an isolated conductor there will be no incoming waves.  For $z>0$ we thus choose a purely right-moving perturbation, $\Psi=\Psi_R(x,y,t-z)$.  From Eqs.~\eqref{chi} and \eqref{eq:pecker} we have $\partial_A \chi = \epsilon_{BA}v^B$ and hence 
\begin{align}\label{treasure}
F = dx \wedge dy + \star v \wedge d(t-z),
\end{align}
where $(\star v)_A=\epsilon_{BA} v^B$  is the two-dimensional dual of the velocity field, evaluated at the retarded time $t-z$.  Eq.~\eqref{treasure} gives the Alfv\'en waves launched by an incompressible conducting flow that begins perturbing a uniform field at some finite time in the past.  It is also an exact solution [the incompressibility  $\nabla_A v^A=0$ guarantees that it can be expressed in the light dart form \eqref{light-dart}], so it is tempting to conclude that it is valid for relativistic velocities.  However, our argument that the fast mode vanishes was only valid in linearized theory, and it cannot be correct nonlinearly.\footnote{Flux conservation dictates that the $z$ component of magnetic field on the conductor must increase at $O(v^2)$ to compensate the Lorentz contraction as the conducting fluid begins to move.  This is not present in Eq.~ \eqref{treasure}, which leaves $B_z$ unperturbed.  Thus while Eq.~\eqref{treasure} may still represent the Alfv\'en waves launched by some relativistic conductor, it is not the same conductor that would have started from rest with field $B_z$ and accelerated up to speed $v$.  We thank Jon McKinney for bringing this point to our attention.}

%This is the \textit{exact} solution for the non-linear Alfv\'en waves launched by an incompressible conducting flow that begins perturbing a uniform field at some finite time in the past.  

The current is given by
\begin{align}\label{current}
j = - \omega  \, (\partial_t + \partial_z),
\end{align}
where $\omega=\epsilon^{AB} \nabla_A v_B$ is the vorticity of the flow.  The Poynting flux in the $z$ direction is given by 
\begin{align}\label{flyingFlux}
 \mathcal{F}_{\rm Poynting} =  v^2,
\end{align}
where $v^2=v^A v_A$ is the flow velocity squared.  For waves launched downward, we would instead use the advanced time $t+z$, with the field and current given by \eqref{treasure} and \eqref{current} (respectively) with $z \rightarrow -z$ and the Poynting flux \eqref{flyingFlux} picking up a minus sign.

Eq.~\eqref{treasure} is remarkably simple and general.  Restoring the overall factor of $B_0$, the electric and magnetic fields are given by  
\begin{subequations}\label{happytimes}
\begin{align}
\vec{E} & = - \vec{v} \times (B_0 \hat{z}) \\
\vec{B} & = B_0( \hat{z} - \vec{v} ),
\end{align}
\end{subequations}
where $v$ is evaluated at the retarded time $t-z$.\footnote{That is, if $\vec{v}(x,y,t)$ is the three-velocity of the flow, then we take $\vec{v}(x,y,t-z)$ in Eqs.~\eqref{happytimes}.}  At each time $t$, the flow velocity pattern is imprinted on the magnetic field and launched out as an Alfv\'en wave traveling at the speed of light.

We may understand the physics of Eqs.~\eqref{happytimes} as follows.  The conducting boundary condition requires $\vec{E}$ to vanish in the rest frame of the fluid, meaning $\vec{E}+\vec{v} \times \vec{B}$ must vanish in the lab frame.  The induced electric field is then ``transmitted'' to the plasma by the junction condition that tangential components of electric fields must be continuous.  The force-free equations then drive a transverse, orthogonal $B$ field and cause the whole disturbance to propagate out at the speed of light.

\section{Jets from Conductors}\label{sec:jets}

If the conducting fluid occupies only a portion of the plane then only that portion will launch Alfv\'en waves, which are then interpreted as a jet.  We describe the boundary by a function $f(x,y,t)=0$,
\begin{align}\label{conductor region}
\textrm{conductor region:} \quad \ z=0, \quad f(x,y,t)<0,
\end{align}
which also defines the jet region by
\begin{align}
\textrm{jet region:} \quad f(x,y,t-z)<0.
\end{align}
The conductor launches waves in both directions, but here we consider $z>0$, with the $z<0$ fields to be determined by reflection about $z=0$.  For simplicity we consider incompressible fluids, so that there are no fast-mode waves.  In this case the field can be described by the light dart form \eqref{light-dart},
\begin{align}\label{clight-dart}
F = dx \wedge dy + d\chi \wedge (dt-dz),
\end{align}
where $\chi$ takes the form
\begin{align}
\chi=\begin{cases}\chi_{\rm in}(x,y,t-z) & f(x,y,t-z)<0 \\ \chi_{\rm out}(x,y,t-z) & f(x,y,t-z)>0 \end{cases}.
\end{align}
The arguments leading to Eq.~\eqref{treasure} are valid in the jet region, meaning $\chi_{\rm in}$ is a stream function for the velocity field,
\begin{align}\label{chiforjet}
\partial_A\chi_{\rm in} = \epsilon_{BA} v^B,
\end{align}
with the velocity field evaluated at time $t-z$.  This results in bulk current flow given by the vorticity $\omega$ at the retarded time $t-z$ [Eq.~\eqref{current}],
\begin{align}
j_{\rm in} = -\omega \, (\partial_t + \partial_z).
\end{align}
The pullback of the field to the jet boundary $f(x,y,t-z)=0$ must be continuous to avoid magnetic monopoles.  This corresponds to the statement that the Wronskian $f_x \chi_y-f_y \chi_x$ must be continuous.  In terms of the normal and tangential vectors  $n_A=\partial_A f/\sqrt{\partial_B f \partial^Bf}$ and $T^A=\epsilon^{BA}N_B$, 
\begin{align}\label{findchi}
T^A \partial_A \chi_{\rm out} = v^A n_A,
\end{align}
holding on the boundary $f(x,y,t-z)=0$.  This is effectively a Dirichlet condition for $\chi_{\rm out}$, since we may integrate \eqref{findchi} along the edge (at fixed $t-z$) to get $\chi_{\rm out}$ everywhere on the boundary.  The result is unique up to an overall constant, which is a gauge freedom of $\chi$.

There still remains an enormous freedom in the choice of $\chi_{\rm out}(x,y,t-z)$.  The ambiguity arises because we have not yet specified boundary conditions on $z=0$ away from the conductor.  To fix the freedom we wish to find an assumption that captures the idea that no physical bodies are present there (besides the charge-carriers in the plasma).  The natural choice is to take the bulk current to vanish [see Eqs.~\eqref{chiforjet} and \eqref{current}],
\begin{align}\label{trichi}
\triangle \chi_{\rm out} = 0,
\end{align}
so that only the physical conductor produces current outflow.  The charge also vanishes, making the entire configuration outside a vacuum solution to Maxwell's equations.  

A final condition is that $\chi_{\rm out}$ must approach a constant as $x$ and $y$ approach infinity, so that the transverse fields fall off and the conductor can be considered isolated.  Along with \eqref{findchi} and \eqref{trichi}, this constitutes a Dirichlet problem in the $2D$ plane minus the conductor region.  Thus $\chi_{\rm out}$ is determined up to the gauge choice of the overall constant, and the entire physical configuration is uniquely determined.

\subsection{Current sheets}

The fields for $z<0$ are determined by reflection $z \rightarrow -z$ of the $z>0$ fields considered above.  In general this entails a current sheet on the $xy$ plane, whose surface current two-form $K$ is given by the jump in the pullback of the dual of the field strength \cite{gralla-jacobson2015}.  Defining the jump as above minus below, from \eqref{light-dart} we have
\begin{align}
K_{\rm plane} = 2 \star \! d \chi \wedge dt,
\end{align}
where $\star$ is dual on the $xy$ plane and the right-hand side is evaluated at $z=0$.    This sheet covers both the conductor region $f(x,y,t)<0$, where $K_{\rm plane} = 2 v \wedge dt$, as well as the outside, where $K_{\rm plane} = 2 \star d \chi_{\rm out} \wedge dt$.  The conductor portion represents surface-current flowing on the top and bottom of the conductor, $j_{\rm   surf} = 2  \delta (z) \, \star \!v$,   while the outside portion is an artifact of considering an infinitesimally thin conductor.  For a finite-thickness conductor, our description should be valid at distances large compared to the thickness.  At comparable distances, the description will be more complicated as the upward-moving jet and associated fields ``transition'' to the downward-moving jet and fields in a continuous manner depending on the detailed shape of the conductor.  Our assumption of a thin conductor compresses this transition region into a sheet on the $z=0$ plane.

Our model also in general contains a current sheet surrounding the jet at $f(x,y,t-z)=0$,
\begin{align}\label{Kjet}
K_{\rm jet} = \Big(  \star  d \chi_{\rm out} + v  \Big)\big\vert_{f=0} \wedge ( dt - dz )
\end{align}
This sheet, by contrast, is ``real'' in the sense that its thickness for a real conductor is set by size of the transition region from being inside the conductor to being outside.  This region will be very narrow for solid body conductors, and the jets launched will be surrounded by a correspondingly narrow region where the charge and current density is large compared with the bulk of the jet.

\subsection{Power radiated}

 The energy per unit time carried away in Poynting flux is %upwards at $z>0$ is given by
\begin{align}\label{powah}
P=2\int \pd_A \chi \pd^A \chi \,dx dy, % = - \int \chi \triangle \chi dx dy, 
\end{align}
where the integral is at any fixed $t$ and $z>0$, with the factor of $2$ accounting for the equal flux leaving downward.  Inside the jet $\partial_A \chi \partial^A\chi =v^2$, but in general there is a contribution from outside as well. 

We can derive an alternative form by choosing the gauge where $\chi$ vanishes at infinity, making $\chi$ the electric potential for the configuration [see Eq.~\eqref{clight-dart}].  The Laplace equation \eqref{trichi} ensures that $\chi$ falls off like inverse cylindrical radius, meaning that the boundary term at infinity will vanish if we integrate \eqref{powah} by parts.  However, there is still a boundary term at the jet edge,
\begin{align}\label{waitforit}
P=-2 \left\{ \int \chi \triangle \chi\, dx dy + \int \chi n^A [\partial_A \chi] ds \right\}.
\end{align}
The second integral is taken on the jet boundary $f=0$ with outward normal $n^A$, and $[\partial_A \chi]=\partial_A \chi_{\rm out}-\partial_A \chi_{\rm in}$ is the jump in the gradient of $\chi$. (The field $\chi$ itself is continuous.)  Since $\triangle \chi$ is the bulk current density and $n^A[\partial_A\chi]$ is the surface current density on the jet edge,  Eq.~\eqref{waitforit} displays the power as the cross-sectional integral of the current density times the electric potential, a generalization of the classic circuit equation $P=IV$.  The factor of $2$ accounts for the presence of two circuits.

In general the current vanishes outside the jet, Eq.~\eqref{trichi}.  In the special case of a vorticity-free source (e.g. the rigidly moving conductors considered in Sec.~\ref{sec:disk} below), Eq.~\eqref{current} shows that the current vanishes inside the jet as well.   In this case the first term is entirely absent and the entire energy flux may be written as the line integral 
\begin{align}\label{Pline}
P = -2\int \Phi_e \ \! J_{\rm surf} \ \! ds.
\end{align}
where $\Phi_e=\chi$ and $J_{\rm surf}=n^A [\partial_A \chi]$ are the electric potential and surface current density on the edge of the jet, respectively.  Since $\Phi_e \propto B v L$ and $J_{\rm surf} \propto B v$, this establishes the basic $B^2 v^2 L^2$ scaling quoted in the introduction.

\section{Disk-shaped conductor}\label{sec:disk}

The general treatment of the previous section gives the field inside the jet in closed form, but requires solution of the Laplace equation to determine the the field outside and the power radiated.  To provide a complete physical model we specialize to the case of a disk (circular) shaped conductor, for which solution of the Laplace equation is straightforward using separation of variables.  We consider a rotating disk, a uniformly moving disk, and a disk moving rigidly with arbitrary velocity.

\subsection{Rotating disk}

Consider a conducting disk of radius $R$ rigidly rotating with angular velocity $\Omega$ about the origin in the $xy$ plane.  We work in cylindrical coordinates $(t,z,\rho,\phi)$.  The conducting surface four-velocity is $u = \gamma(\partial_t + \Omega \partial_\phi)$ with $\gamma \equiv (1-\Omega^2 R^2)^{-1/2}$ and its region is described by $f<0$ with $f=\rho^2-R^2$.  The flow is incompressible but has vorticity $\omega = 2 \Omega$.  The stream function in the jet region $\rho<R$ is $\chi_{\rm in} = -\tfrac{1}{2} \rho^2 \Omega$.
%\begin{equation}\label{Psirot}
%\chi_{\rm in} = \tfrac{1}{2} \rho^2 \Omega,
%\end{equation}
%holding in the jet region $f<0$,
%\begin{align}
%\textrm{jet region: } \quad \rho<R.
%\end{align}
Since the flow is tangent to the edge $\rho=R$, Eq.~\eqref{findchi} says that $\chi_{\rm out}$ must be constant there.  The unique solution to $\triangle \chi_{\rm out}=0$ vanishing at infinity is just $\chi_{\rm out}=0$.  Including the region $z<0$ by reflection symmetry and restoring the overall factor of $B_0$, the final solution is given by
\begin{equation}\label{Frot}
F = B_0 \rho \,d\rho \wedge \Big[ d \phi - \Omega (dt - \epsilon(z) dz) H(R-\rho) \Big] , 
\end{equation}
where $H(x)$ is the Heaviside function ($1$ for positive $x$ and $0$ for negative $x$), while $\epsilon(z)$ is defined to be the sign of $z$. The total power radiated per unit time (including both ``jets'') is given by 
\begin{align}\label{Prot}
P =  \pi B_0^2 \Omega^2 R^4.
\end{align}
Note that Eqs.~\eqref{Frot} and \eqref{Prot} remain valid for time-dependent rotation velocity $\Omega(t)$ with the substitution $\Omega\rightarrow \Omega(t-z)$. 

\begin{figure}
\includegraphics[width=.45\textwidth]{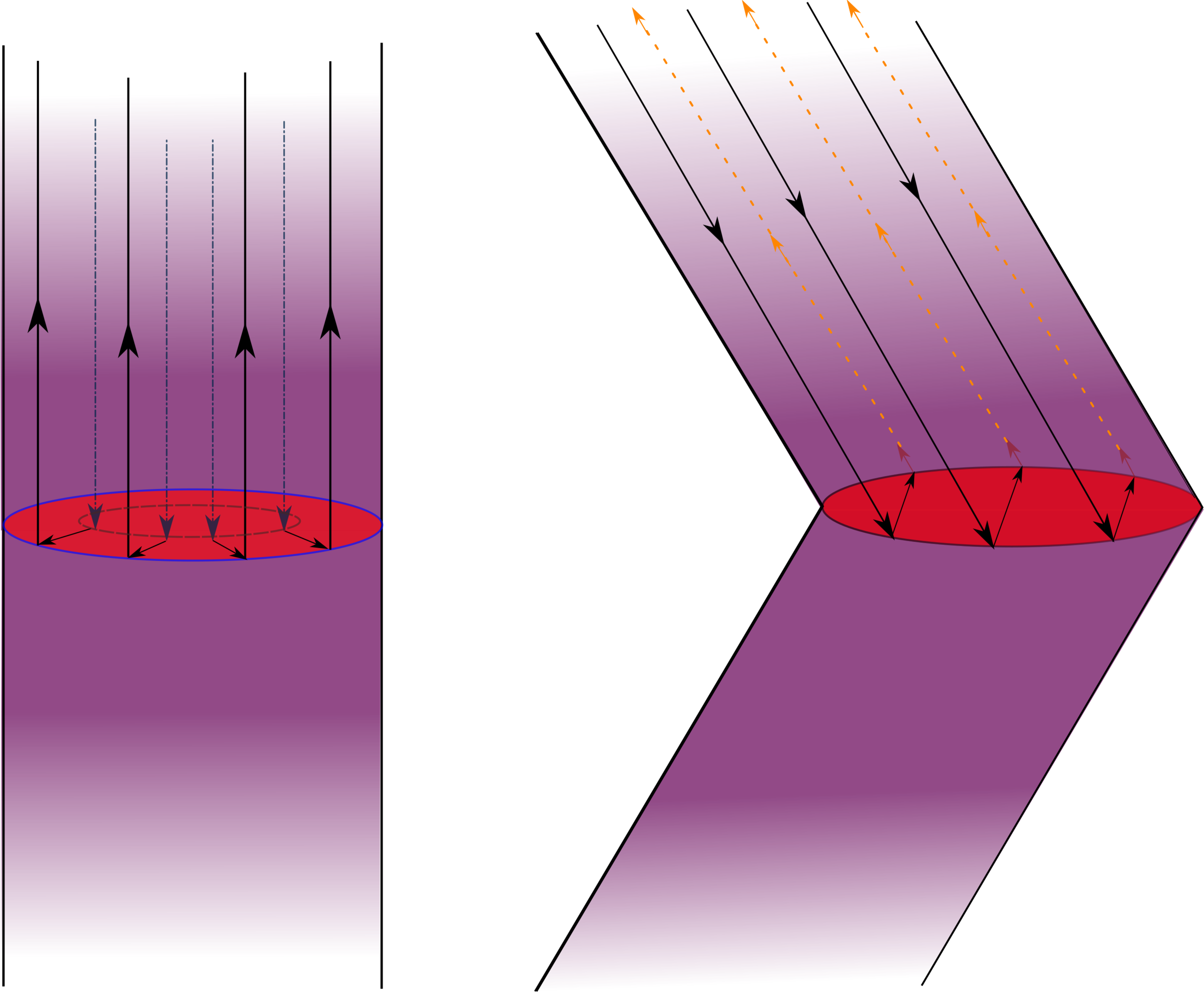}
\caption{Current flow in the jet: comparison of rotating (left) and moving (right) cases.  For the moving case, we show the rest frame current, which is tangent to the jet.  In the rotating case there is both bulk and surface current flow, while in the moving case there is only surface flow.}  \label{fig:current}
\end{figure} 

The charge-current naturally splits into three pieces,
\begin{align}
j =  j_{\rm bulk} + j_{\rm edge} + j_{\rm disk}
\end{align}
where
\begin{subequations}\label{jrot}
\begin{align}
j_{\rm bulk} & = -2 B_0 \Omega \big(\pd_t  + \epsilon(z)  \pd_z \big) \ \!  H(R-\rho)  \\
j_{\rm edge} & = B_0 \rho \Omega   \big(   \pd_t + \epsilon(z) \pd_z \big)  \ \! \delta(\rho-R) \quad  \\
j_{\rm disk} & =  2  B_0 \rho  \Omega   \, \delta(z)  \ \! H(R-\rho) \pd_\rho  .
\end{align}
\end{subequations}
Bulk current flows inside the jet(s) with equal and opposite surface current on the boundary.  Surface current flowing on the disk closes the circuit in the equatorial plane, while the circuit is ``open'' as $z \rightarrow \pm \infty$.  This structure is illustrated in Fig.~\ref{fig:current}.  In the language of steady state circuits, the current $I$ can be computed from any of the legs \eqref{jrot} of the circuit, while the voltage $V$ can be computed from the edge to the center.  We can then ascribe an effective resistance $\mathfrak{R} := \frac{V}{I}$ to the circuit.  These quantities are given by
\begin{align}
I =  B_0 R^2 \Omega, \ \ \ V = \tfrac 12 \, B_0 R^2  \Omega, \ \ \ \mathfrak{R} = \frac{1}{2}.
\end{align}

\subsection{Moving Disk} 

Now suppose that the thin conducting disk is moving in the $x$ direction at constant velocity $v$.  Then the conductor region is $z=0,f<0$ with $f=(x-vt)^2 +y^2 - R^2$ and the four-velocity field is given by $u=\gamma \left(\pd_t + v \pd_x \right)$ on that region.\footnote{This assumes a circular shape in the lab frame.  More physically, we should assume a circular shape in the rest frame, which would give rise to an elliptical shape in the lab frame.  This difference is irrelevant at leading order in $v$.}  The flow is incompressible and the jet region is given by
\begin{align}
\textrm{jet region:} \quad \xi^2 + y^2 < R^2,
\end{align}
where
\begin{align}
\xi := x-v(t-z).
\end{align}
Thus the jet is bent backwards by an angle $\theta_{\rm jet} = \tan^{-1} v$.  This is a general feature of jets from moving conductors, not limited to the circular shape considered here.  The fields inside the jet are given by the stream function for the uniform flow,
\begin{align}\label{chiin1}
\chi_{\rm in} = v y.
\end{align}
Working in cylindrical coordinates $\rho,\phi$ defined relative to $\xi$ and $y$, the boundary condition \eqref{findchi} becomes $\partial_\phi \chi_{\rm out}=v \cos \phi$.  Separating variables, the unique solution to the Laplace equation \eqref{trichi} vanishing at infinity is seen to be
\begin{align}\label{chiout1}
\chi_{\rm out} = \frac{v y}{\xi^2+y^2} \, R^2,
\end{align}
the dipolar solution regular at infinity.  The field strength is reconstructed by \eqref{light-dart} using \eqref{chiin1} and \eqref{chiout1} on the inside and outside of $\xi^2+y^2=R^2$, respectively.  (Additionally one must add on the $z<0$ fields by reflection $z \rightarrow -z$.)  Restoring the overall factor of $B_0$, the total energy per unit time flowing away from the moving conductor is 
\begin{align}\label{Pmove}
P = 4 \pi v^2 R^2 B_0^2.
\end{align}

We divide up the current in analogy with the rotating case,
\begin{align}
j = j_{\rm bulk} + j_{\rm edge} + j_{\rm disk} + j_{\rm plane}
\end{align}
where here
\begin{subequations}\label{jmove}
\begin{align}
j_{\rm bulk} & = 0 \\
j_{\rm edge} & = 2 v B_0 y \, \delta( \sqrt{\xi^2+y^2} -R ) \,   \left(\pd_t + \epsilon(z) \pd_z \right) \quad  \\
j_{\rm disk} & = 2 v B_0 H(R-\sqrt{\xi^2+y^2})\delta(z) \pd_y\\\
j_{\rm plane} &= \frac{ 2 v B_0}{(\xi^2 + y^2)^2}\Big( 2 y \xi \pd_x + (y-\xi)(y+\xi) \pd_y \Big) \nonumber \\& \quad\,\,\,\,  \times H(\sqrt{\xi^2+y^2} -R) \delta (z).
\end{align}
\end{subequations}
We have restored the overall factor of $B_0$.  Notice that there is \textit{no} current flowing in the interior of the jet, a consequence of the vanishing vorticity of the source.  In addition to the main circuit, there is also circulating current in the plane outside the source.

%Notice that there is \textit{no} current flowing in the interior of the jet---all of the current flows in the thin layer on the edge (and on the disk).  This corresponds to the fact that the uniform conducting flow has vanishing vorticity.

As in the rotating case, the current $I$ may be computed from any of the legs, the voltage $V$ may be computed from edge to edge across the jet, and we may ascribe an effective resistance $\mathfrak{R}=V/I$.  These quantities are given by
\begin{align}
I = 4 R v B_0, \quad V = 2 R v B_0, \quad \mathfrak{R} = \tfrac{1}{2}.
\end{align}

\subsubsection{Comoving frame}

In the rest (comoving) frame the fields are stationary and the electric field is zero inside the jet. Denoting this frame with a prime, the fields inside the jet are\footnote{The shape of the boundary between the inside and outside also changes under the boost.  This change is $O(v^2)$ and can be ignored to leading order.}
\begin{subequations}
\begin{align}
\vec{E'}_{\rm in} &=  0 , \\
\vec{B'}_{\rm in} &= \gamma (\hat{z}' - v \hat{x}')B_0, 
\end{align}
\end{subequations}
while those outside are
\begin{subequations}
\begin{align}\label{Eout} 
\vec{E'}_{\rm out} &= \vec{\nabla}' \Xi \\
\vec{B'}_{\rm out} &= B_0 \hat{z}' + \hat{z}' \times \vec{\nabla}' \Xi,
\end{align}
\end{subequations}
where the rest frame electric potential $\Xi$ is given by
\begin{align}\label{Xi}
\Xi = \gamma \left( \chi_{\rm out} - y' v  \right) B_0 .
\end{align} 
From \eqref{Eout} and \eqref{Xi}, we see that the boost generates a asymptotic electric field in the $\hat{y}'$ direction. The three-current flowing on the edge of the jet in the rest frame is given by
\begin{align}\label{Jprime}
\vec{J}' = 2 v B_0 y' \Big(\epsilon(z')\hat{z}' - \gamma v \hat{x}' \Big) \delta \left( \sqrt{\xi^2+y^2} - R^2 \right),
\end{align}
which is tangent to the jet edge, as illustrated in Fig.~\ref{fig:current}.

The level sets of $\Xi$ have the dual interpretations of equipotential surfaces for the electric field and integral curves of the comoving drift velocity ${\vec{v}'}_D=\vec{E}' \times \vec{B}'/\vec{B}'\cdot\vec{B}'$.
The latter represents the velocity field of plasma particles that are at rest in the lab frame far from the conductor.\footnote{In the force-free approximation, the only knowledge retained of the underlying plasma particles is that they move along on the magnetic field sheets \cite{gralla-jacobson2014}, which corresponds to motion along magnetic field lines defined in any orthonormal frame where the electric field is everywhere vanishing.  The actual motion is dependent upon initial conditions for motion along each field line, with the drift velocity of each Lorentz frame corresponding to one choice.  The lab-frame drift velocity ${\vec{v}'}_D$ approaches (minus) the conductor velocity far from the jet, and hence corresponds to the choice of particles at rest in the magnetic field frame before being disturbed by the conductor.}  Thus if the conductor moves through an otherwise undisturbed plasma, it finds itself bathed in a wind of particles following the streamlines of ${\vec{v}'}_D$.  This wind is plotted in Fig.~\ref{fig:streamdisk} to linear order in $v$. 

\subsubsection{Ideal flow and other shapes}

For the disk-shaped conductor we found the solution in the lab frame and then transformed to the rest frame.  For conductors of other shapes it is convenient to think directly in the rest frame.  To linear order in $v$, the equations are the Laplace equation $(\partial_\xi^2+\partial_y^2)\Xi=0$, the Dirichlet condition $\Xi = \rm const$ on the conductor edge (corresponding to the vanishing of the tangential components of the electric field), and the asymptotic behavior $\Xi \rightarrow - v y B_0$ at large $\xi$ and $y$.  These are the equations of a two-dimensional ideal fluid flow past an obstacle, a classic problem that can be solved by conformal maps.

\begin{figure}
\subfigure[\ circle]{\label{fig:streamdisk}\includegraphics[height=0.21\textwidth]{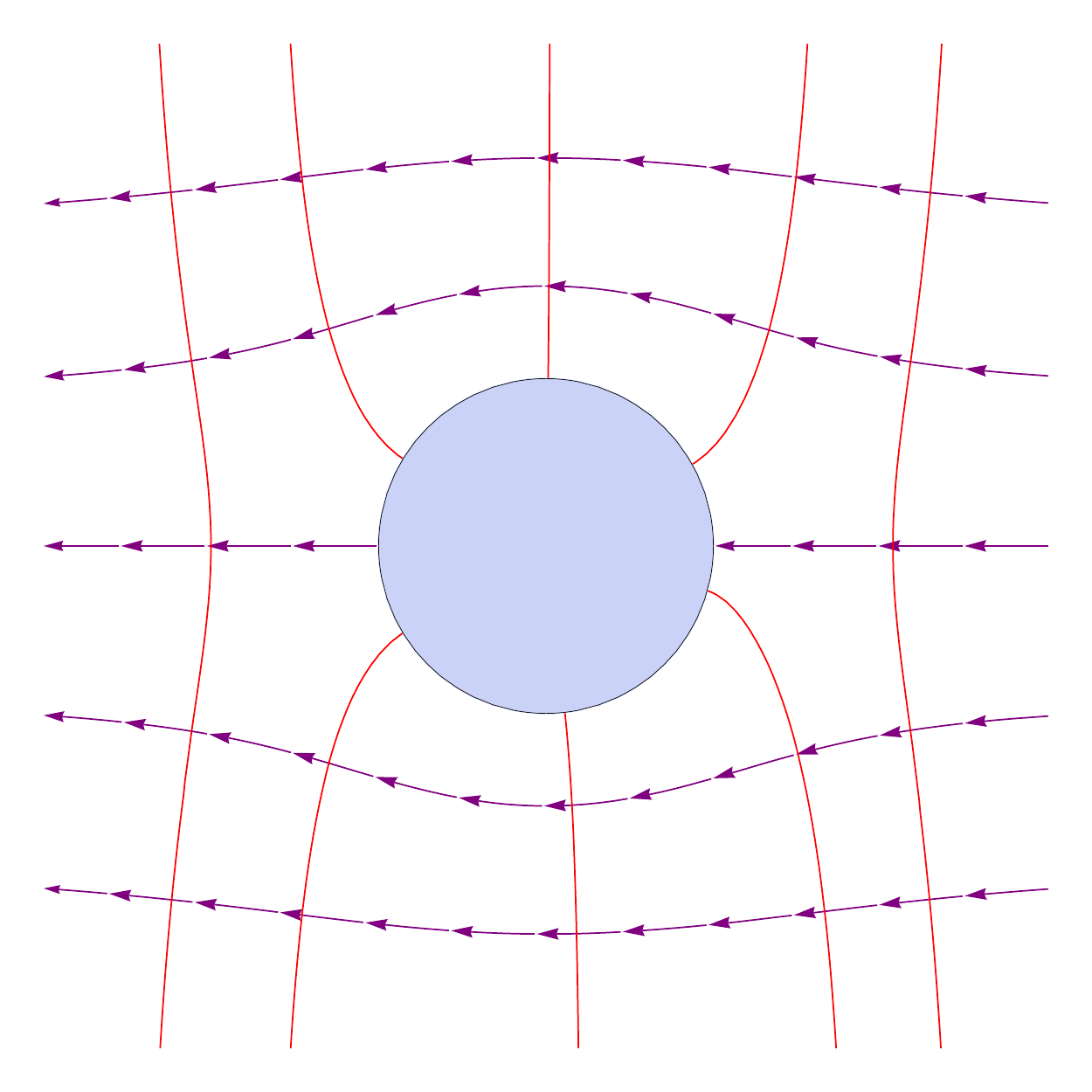} \ \ \ \ \ \  }
\subfigure[\ ellipse]{\label{fig:streamellipse}\includegraphics[height=0.21\textwidth]{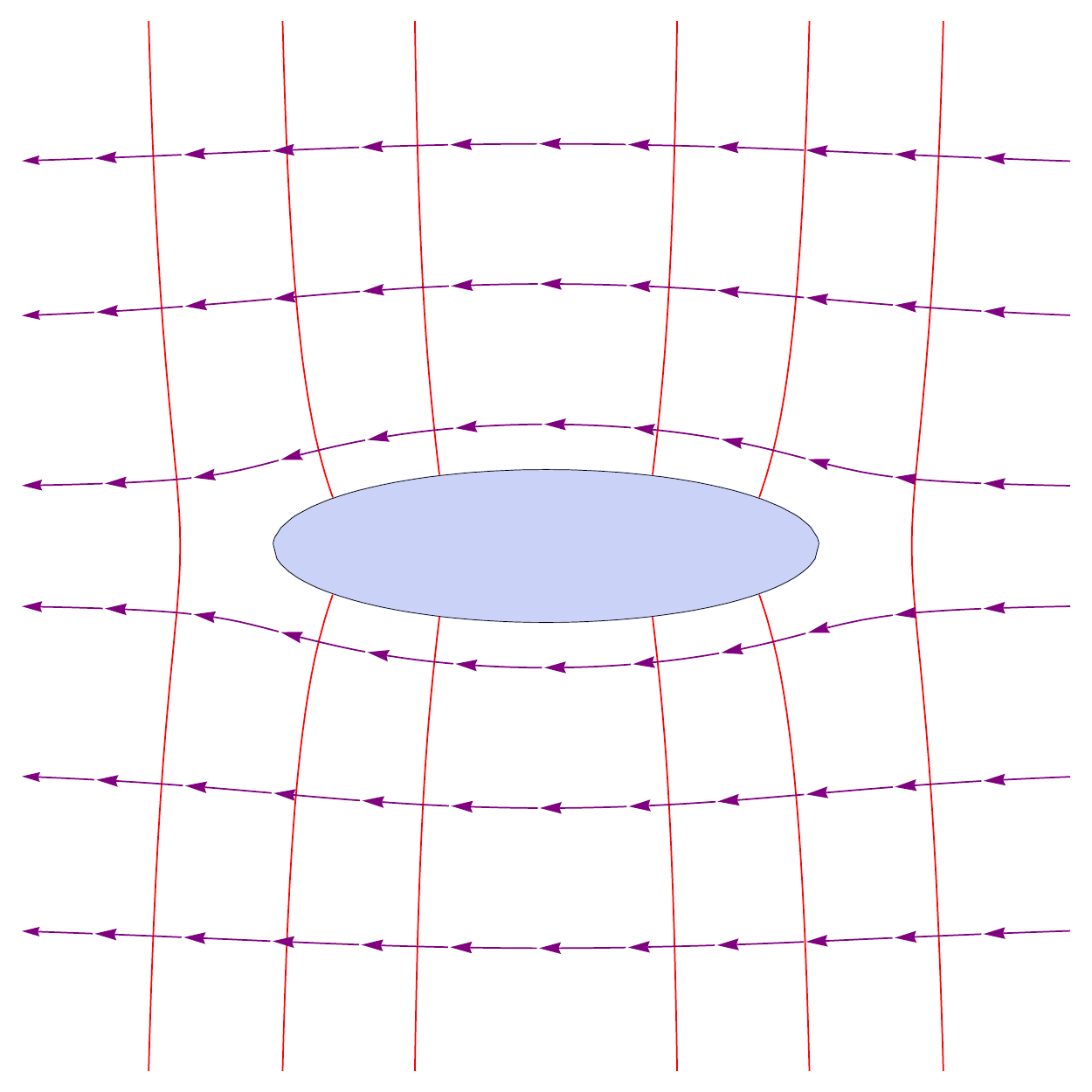} }
\caption{Top-down view of the plane of the moving conductor, showing rest-frame electric field lines in red and plasma flow (drift velocity) in purple.  We show the circular disk as well as the solution for an elliptical disk, found by a conformal map. The plasma flow is identical to an ideal fluid flow past an obstacle.}\label{fig:fluid}
\end{figure} 

In particular, having the solution \eqref{Xi} for the circular conductor, we can apply any conformal map preserving the asymptotic behavior and obtain the solution for the motion of a conductor whose shape is the image of the disk under the map.  In particular, note that $\Xi$ \eqref{Xi} and its harmonic conjugate are the real and imaginary parts of the anti-holomorphic\footnote{ The potential is anti-holomorphic, $\pd_Z U =0$, because it generates a leftward flow.} function $U(\bar Z) = \bar Z + \frac{1}{\bar Z}$, where $\bar Z = \xi - i y$.  New solutions are generated by mapping $\bar Z$ to a new function $W(\bar Z)$.  In Fig. \ref{fig:streamellipse}, we display the flow velocity and the electric field lines for the ``ellipse'' mapping $W(\bar Z) = \bar Z + \varepsilon^2 /\bar{Z}$, where $\varepsilon$ is the ellipticity parameter $\varepsilon \leq 1$.  

\subsection{Non-uniform Motion}

Finally suppose that the disk moves rigidly along an arbitrary trajectory $x^A=X^A(t)$ with velocity $V^A=dX^A/dt$. It is convenient to introduce $\xi^A$ defined by
\begin{align}
\xi^A := x^A -X^A( t - z ),
\end{align}
holding in Cartesian coordinates $\{x,y\}$ only.  Following the same steps as in the uniform motion case, we find
\begin{subequations}\label{nonuniform}
\begin{align}
\textrm{jet region:} &\quad  \delta_{AB} \xi^A \xi^B < R^2, \\
\chi_{\rm in} &=  \epsilon_{AB}V^A \xi^B  \label{chiin} \\
\chi_{\rm out} &= \frac{\epsilon_{AB} V^A \xi^B  }{\delta_{AB} \xi^A \xi^B} \, R^2,\label{chiout}
\end{align}
\end{subequations}
where $V^A$ is evaluated at time $t-z$.

\subsection{Jets from a binary}
In pioneering work on dynamical force-free fields in General Relativity, Refs.~\cite{palenzuela-etal2010,palenzuela-lehner-liebling2010} presented numerical simulations of binary black holes embedded in a uniform magnetic field and force-free plasma.  This work revealed a characteristic braided dual-jet structure that was subsequently confirmed in \cite{alic-etal2012}.  One can expect a similar structure if two orbiting conductors (such as neutron stars) are immersed in a magnetic field, and, noting that superposition is valid to linear order in the velocity, we are now in a position to present corresponding analytic solutions.   Using \eqref{nonuniform} for two trajectories that orbit each other produces the analytic solution for jets from a binary, proudly displayed in Fig~\ref{fig:binaryjet}.  

\section*{Acknowledgements}
We thank Maxim Lyutikov, Jon McKinney, Kyle Parfrey, and Leo Stein for helpful conversations.  This work was supported in part by NSF grant PHY--1506027 to the University of Arizona.
\bibliography{plasmawaves}
\end{document}